\documentclass[submission,copyright,creativecommons]{eptcs}
\providecommand{\event}{SYNT 2015} 

\usepackage{graphicx}
\usepackage{array}
\usepackage{booktabs}
\usepackage{multirow}
\usepackage{tikz}

\title{Results and Analysis of SyGuS-Comp'15}
\author{
Rajeev Alur \qquad\quad
Dana Fisman
\institute{University of Pennsylvania}
\and
Rishabh Singh
\institute{Microsoft Research}
\and
Armando Solar-Lezama
\institute{Massachusetts Institute of Technology}
}
\def\titlerunning{Results and Analysis of SyGuS-Comp'15}
\def\authorrunning{R. Alur, D. Fisman, R. Singh \& A. Solar-Lezama}
\begin{document}
\maketitle

\newcommand{\alc}{\textsc{Alchemist-cs}}
\newcommand{\alccsdt}{\textsc{Alchemist-csdt}}
\newcommand{\cvc}{\textsc{Cvc4-1.5-sygus}}
\newcommand{\enum}{\textsc{Enumerative}}
\newcommand{\skac}{\textsc{Sketch-ac}}
\newcommand{\ice}{\textsc{Ice-dt}}
\newcommand{\toast}{\textsc{SosyToast}}
\newcommand{\stoch}{\textsc{Stochastic}}
\newcommand{\sygus}{SyGuS}
\newcommand{\comp}{SyGuS-Comp}

\begin{abstract}
\emph{Syntax-Guided Synthesis (SyGuS)} is
the computational problem of finding an implementation $f$ that
meets both a semantic constraint
given by a logical formula $\varphi$ in a background theory $T$,
and a syntactic constraint given by a grammar $G$, which specifies the allowed set of
candidate implementations.
Such a synthesis problem can be formally defined in SyGuS-IF,
a language that is built on top of SMT-LIB.

The \emph{Syntax-Guided Synthesis Competition (\comp)} is an
effort to facilitate, bring together and accelerate research and development of efficient
solvers for SyGuS by providing a platform for evaluating different synthesis
techniques on a comprehensive set of benchmarks. In this year's competition we added two specialized tracks: a track for conditional linear arithmetic, where the grammar need not be specified and is implicitly assumed to be that of the LIA logic of SMT-LIB, and a track for invariant synthesis problems, with special constructs conforming to the structure of an invariant synthesis problem. This paper presents and analyzes the results of \comp'15.
\end{abstract}

\section{Introduction}
\label{sec:intro}
Syntax guided synthesis is a form of synthesis where in addition to a specification, the user provides a syntactic description of the space of possible programs that the synthesizer is allowed to consider. This style of synthesis was popularized by the Sketch project~\cite{solar2006combinatorial} and has become an important approach underpinning a variety of synthesis efforts (e.g. ~\cite{Jha2010, SrivPts, TorlakB14}). Within the common framework of syntax guided synthesis, however, there are a number of possible approaches that can be used to solve a given synthesis problem, and until recently, research on the relative merits of these different approaches had been hampered by the difficulty of making direct comparisons about algorithms implemented in different systems and with very different interfaces. 

The Syntax Guided Synthesis competition (\comp{}) was started as a way to accelerate advances in the field by offering a uniform standard notation and formalism for describing these problems, allowing for direct comparisons between different solution methods.  The \sygus{} formalism~\cite{RaghothamanU14} is derived from the SMT-LIB format~\cite{smtlib}, a popular format for describing SMT problems, with the goal of making it easy to learn for users and easy to support for synthesizer developers. In the short time that the formalism has been in public circulation, it has already performed well in many of these goals. It has provided important insights into the relative merits of different algorithms~\cite{AlurBDF0JKMMRSSSSTU15,AlurBJMRSSSTU13} which have been exploited to help develop new algorithms~\cite{JeonQSF15}, and has been used to evaluate brand new algorithms~\cite{AlurCAV15,AleksICSE15, ReynoldsDKTB15}. Beyond synthesizer developers, there is a growing community of users that is coalescing around the formalism.

The \sygus{} problem is formally defined as the problem of finding an implementation of a function $f$ that meets both a semantic constraint given by a logical formula $\varphi$ in a background theory $T$, and a syntactic constraint given by a grammar G. More concretely, let $\varphi$ be a formula in some theory $T$ that involves some free variables $x_0, \ldots x_i$ and uses an unknown function $f$. The goal of synthesis is to discover an implementation $f_{imp}$ such that replacing $f_{imp}$ for $f$ in $\varphi$ makes the formula valid for all values of the free variables. 
\[
\forall x_0, \ldots x_i. ~ \varphi[f/f_{imp}]
\]
The syntactic restriction implies that the function $f_{imp}$ must belong to the grammar $G$ also provided by the user.  

\paragraph{Example}
As an example, consider the problem of finding a bit-vector function that returns a bit-vector with a $1$ in the position of the least significant zero in its input bit-vector and zero everywhere else. This problem is defined in terms of the theory of bit-vectors, which includes operations such as bitwise \texttt{and}, bitwise \texttt{or}, as well as left and right shifts and basic arithmetic among bit-vectors. 
The example problem can be formalized in terms of two constraints that must be satisfied by the desired function. The first constraint is that if $x$ is not zero, $f(x)$ should have a one in the same position as $x$ has a zero, so bitwise \texttt{and} (\texttt{\&}) of $f(x)$ and the bitwise negation of $x$ ($\sim$\texttt{x}) should not be zero.
\[
\forall x.~ x > 0 \Rightarrow (f(x)~ \mbox{\texttt{\&}}~ \sim x)  > 0  
\]
The way we enforce that there is a single one and that it corresponds to the least significant zero is by enforcing that if we shift $f(x)$ right by any amount, then bitwise anding with the bitwise negation of $x$ will now yield zero.
\[
\forall x, y .  ~ y> 0 \Rightarrow (f(x) \gg y~ \mbox{\texttt{\&}}~ \sim x) =0 
\]

These two constraints can be expressed succinctly in \sygus\ with the following notation.
\begin{verbatim}
(set-logic BV)

(synth-fun f ((x (BitVec 32))) (BitVec 32)
             ((Start (BitVec 32) (x 0 1 
                                 (bvand Start Start)  
                                 (bvor Start Start)
                                 (bvnot Start) 
                                 (bvadd Start Start)))))

(declare-var x (BitVec 32))
(declare-var y (BitVec 32))

(constraint (=> (bvult 0 x) (bvult 0 (bvand (f x) (bvnot x)))))
(constraint (=> (bvult 0 y) (= 0 (bvand (bvshr (f x) y) (bvnot x)))))

(check-synth)
\end{verbatim}
The \texttt{set-logic} directive indicates that the constraints should be interpreted in terms of the theory of bitvectors. The directive \texttt{declare-var} is used to declare \texttt{x} and \texttt{y} as 32-bit bitvector variables. The constraints are introduced with the directive \texttt{constraint}, and \texttt{check-synth} marks the end of the problem and prompts the synthesizer to solve for the missing function.
Crucially, in order for the synthesizer to generate $f$, it needs a grammar, which is provided as part of the \texttt{synth-fun} directive.
In this example, we have significantly reduced the space of possible functions by restricting the search to expressions involving bitwise \texttt{and}, \texttt{or}, bitwise negation, sum and the constants 0 and 1, instead of asking the system to consider completely arbitrary expressions in the theory of bitvectors.

\subsection{Conditional Linear Integer Arithmetic Track}
For problems where the grammar consists of the set of all possible integer linear arithmetic terms, it is sometimes possible to apply specialized solution techniques that exploit the information that decision procedures for integer linear arithmetic are able to produce. For this reason, the 2015 \sygus{} competition included a separate track where the grammar for all the unknown functions was assumed to be the entire theory of Integer Linear Arithmetic with ITE conditionals.

\paragraph{Example} 
As a simple example, consider the problem of synthesizing a function \texttt{max2} that produces the maximum of two integers. The problem can be specified with the constraint below:
\begin{verbatim}
(set-logic LIA)
(synth-fun max2 ((x Int) (y Int)) Int)
(declare-var x Int)
(declare-var y Int)
(constraint (>= (max2 x y) x))
(constraint (>= (max2 x y) y))
(constraint (or (= x (max2 x y)) (or (= y (max2 x y)))))
(check-synth)
\end{verbatim}

Note that the definition of the unknown function \texttt{max2} does not include a grammar this time, but because the problem is defined in the theory of linear integer arithmetic (\texttt{LIA}), the default grammar consists of all the operations available in the theory.

\subsection{Invariant Synthesis Track}
One of the main applications of \sygus{} is invariant synthesis. For this problem, the goal is to discover an invariant that makes the verification condition for a given loop valid. Such a problem can be easily encoded in \sygus{}, but invariant synthesis problems have structure that some solution algorithms are able to exploit and that can be lost when encoding it into \sygus{}. For this reason, the 2015 installment of the competition also included a separate track for invariant synthesis problems where the additional structure is made apparent. In the invariant synthesis version of the \sygus\ format, the constraints are separated into pre-condition, post-condition and transition relation, and the grammar for the unknown invariant is assumed to be the same as that for the conditional linear arithmetic track. 

\paragraph{Example} 
For example, consider the following simple loop. 
\begin{verbatim}
Pre: i >= 0  and  j=j0 and i=i0;
while(i > 0){
    i = i - 1;
    j = j + 1;
}
Post: j = j0 + i0;
\end{verbatim}
Suppose we want to prove that the value of j at the end of the loop equals the value of i + j at the beginning of the loop. The verification condition for this loop would check that (a) the precondition implies the invariant, (b) that the invariant is inductive, so if it holds before an iteration and the loop condition is true, then it will hold after that iteration, and (c) that the invariant together with the negation of the loop condition implies the postcondition. All of these constraints can be expressed in the standard \sygus{} format, but they can be expressed more concisely using the extensions explicitly defined for this purpose. Specifically, the encoding will be as follows.
\begin{verbatim}
(set-logic LIA)

(synth-inv inv-f ((i Int) (j Int) (i0 Int) (j0 Int)))

(declare-primed-var i0 Int)
(declare-primed-var j0 Int)
(declare-primed-var i  Int)
(declare-primed-var j  Int)

(define-fun pre-f ((i Int) (j Int) (i0 Int) (j0 Int)) Bool
                  (and (>= i 0) (and (= i i0) (= j j0))))

(define-fun trans-f ((i Int) (j Int) (i0 Int) (j0 Int)
                     (i! Int) (j! Int) (i0! Int) (j0! Int)) Bool
                     (and (and (= i! (- i 1)) (= j! (+ j 1)))
                          (and (= i0! i0) (= j0! j0))))

(define-fun post-f ((i Int) (j Int) (i0 Int) (j0 Int)) Bool
                   (= j (+ j0 i0)))

(inv-constraint inv-f pre-f trans-f post-f)

(check-synth)
\end{verbatim}

The directive \texttt{(declare-primed-var i)} is equivalent to separately declaring \texttt{i} and \texttt{i!}, where the primed version of the variables is used to distinguish their value before and after the loop body. Just like in the earlier example, the function to be synthesized \texttt{inv\_f} does not include a grammar, so the entire \texttt{LIA} grammar is assumed. Here the return type is also not given and is assumed to be \texttt{Bool} since invariant are assumed to be predicates. The constraint \texttt{inv-constraint} is syntactic sugar for the full verification condition involving the invariant, precondition, postcondition and transition function.

\subsection{\comp{}'14 summary}
The first \sygus\ competition, \comp'14\ consisted of a single track --- the general track --- in which the benchmark provides the grammar describing the desired syntactic restrictions for that benchmark. The background theory could be either linear interger arithmetic or bitvectors. Five solvers competed in \comp'14. The solver who won the first place was the \enum\ solver which solved 126 out of 241 benchmarks.

\subsection{\comp{}'15 summary}

The 2015 instance of  \comp{} was the second iteration of the competition and the first iteration to include the separate conditional linear integer arithmetic and invariant synthesis tracks. As elaborated in Section~\ref{sec:setting}, there were a total of eight solvers submitted to the competition which represent a range of solution strategies. In the rest of the paper, we describe the details of the benchmarks used for the competition, the solution approaches used by each of the participants and the results of the competition on each of the different categories of benchmarks.

\section{Competition Settings}
\label{sec:setting}
\subsection{Participating Benchmarks}
\label{subsec:categories}
We collected several benchmarks from various sources using an open call for benchmarks, from which we sampled $309$ benchmarks for the General Track, $73$ benchmarks for the Invariant Synthesis Track, and $67$ benchmarks for the Conditional Linear Arithmetic track. These benchmarks can broadly be classified into following $13$ categories ($11$ categories for the General Track). The number of benchmarks in each of these categories and the contributors for the new set of benchmarks is shown in Table~\ref{tbl:benchmarks-number}.

\begin{itemize}

\item{\textbf{Array Benchmarks}}: This category consists of benchmarks that involve synthesizing a function over an integer array of a bounded size. Since the current SyGuS-IF format does not support arrays, we represent them using an ordered sequence of integer variables. One major class of benchmarks in this category is \texttt{array-search}, which requires to synthesize a loop-free function to find an appropriate index for insertion of a given value into a sorted array of size $n$.  Another major class of benchmark is \texttt{array-sum-m-n}, which requires to synthesize a function to find first two elements in an array of size $n$ whose sum is greater than $m$.

\item{\textbf{Let Benchmarks}}: The benchmarks in this category use the \texttt{let} construct as specified in the SyGuS-IF format. The let expressions allow synthesizers to represent multiple occurrences of common subexpressions succinctly, which helps in improving the scalability. The let expressions in specification constraints can be desugared, but they can not be desugared when used in grammar productions. For example, the production \texttt{T := (let [z=U] z + z)} denotes sum of two identical subexpressions built using addition operator, but the same can not be specified using a context-free grammar.

\item{\textbf{Invariant Generation Benchmarks (Bounded and Unbounded Integers)}}: This category of benchmarks consists of loop invariant synthesis problems from the domain of program verification. Some of these benchmarks are obtained from the Competition of Software Verification (SV-COMP) and others are from the literature on invariant synthesis. It involves two sub-categories: one that uses unbounded integers and second that bounds the range of integers that can be used for the invariant expression.

\item{\textbf{ICFP Benchmarks}}: This benchmarks in this category are some of the challenging problems taken from the 2013 ICFP Programming Competition on program synthesis. These benchmarks involve synthesizing bit-vector functions from a domain-specific language of bit-vectors consisting of primitive bitwise operators such as shift, not, add, or, xor etc. The language also consists of a constrained comparison operator and a fold operator. The correctness specification for these benchmarks is provided using 64-bit input-output bit-vector examples.

\item{\textbf{Hacker's Delight Benchmarks}}: This category involves benchmarks that are derived from 20 different bit-manipulation problems from the Hacker’s Delight book. For these benchmarks, there are 3 increasingly challenging levels of grammars ($d0$, $d1$, and $d5$) that are provided for the unknown function, where the level $d0$ consists of only operators that are necessary for the unknown function and the level $d5$ consists of complete bit-vector grammar.

\item{\textbf{Integer Benchmarks}}: These benchmarks involve synthesizing functions that involve complex branching structure over linear arithmetic expressions. For example, the \texttt{max} benchmarks require to synthesize a function using comparison operator to compute maximum of $n$ integers.

\item{\textbf{BitVector Benchmarks}}: This category of benchmarks involve synthesizing a complex function over a set of Boolean values represented using a bit-vector. For example, the \texttt{parity} benchmarks compute the parity of a given set of Boolean values, and the \texttt{Morton} benchmarks compute the Morton numbers.

\item{\textbf{Compiler Optimization Benchmarks}}: The benchmarks in this category come from the domain of compiler optimization where the goal is to synthesize a simpler expression (constrained by a grammar) that is functionally equivalent with a given complicated expression.

\item{\textbf{Multiple Functions Benchmarks}}: This class of benchmarks consists of synthesis problems that involve synthesizing multiple unknown functions that satisfy a given set of constraints.

\item{\textbf{Motion Planning Benchmarks}}: The benchmarks in this category come from the domain of robot motion planning, where the task is to synthesize a function to control robot movement (constrained by a grammar of possible movements) that reach from one point in space to another while maintaining some invariant constraints such as no collision.
 
\item{\textbf{LIA Track Benchmarks}}: The benchmarks in this category come from the General Track categories such as integer benchmarks, array benchmarks, and motion planning benchmarks. The grammar for the unknown functions for this category comprises of expressions from the entire theory of Integer Linear Arithmetic with ITE conditionals as described in Section~\ref{subsec:bench-features}.

\item{\textbf{INV Track Benchmarks}}: The benchmarks in this category come from the domain of program verification, where the pre-condition, transition function, and the post-condition are specified using explicit constructs in the SyGuS-IF format.
 
\end{itemize}

\begin{table}[]
\centering
\begin{tabular}{lcrr}
\hline
\hline
\textbf{Benchmark Cateogry}      & \textbf{Number of Benchmarks} & \textbf{Contributors (in 2015)}     \\
\hline
\hline
Arrays                  & 31                   &                          &  \\ \hline
Bitvectors              & 5                    &                          &  \\ \hline
CompilerOptimizations   & 21                   & Nissim Ofek (Yale):        & 21  \\ \hline
HackersDelight          & 44                   &                          &  \\ \hline
ICFP                    & 50                   &                          &  \\ \hline
Integers                & 34                   & Shambwaditya Saha (UIUC):  &  14 \\ \hline
InvariantGeneration     & 28                   &                          &  \\ \hline
InvariantGenerationUbdd & 28                   &                          &  \\ \hline
Let                     & 15                   &                          &  \\ \hline
MotionPlanning          & 12                   & Sarah Chasins (UC Berkeley): & 12 \\ \hline
MultipleFunctions       & 41                   &                           & \\ \hline
INV Track               & 73                   & Pranav Garg (UIUC):         & 51 \\ \hline
LIA Track               & 67                   &                           & \\ \hline
\end{tabular}
\caption{The number of benchmarks in each individual category used in SyGuS-COMP 2015 together with the number of new contributions and their contributors.}\label{tbl:benchmarks-number}
\end{table}

\subsection{Participating Solvers}
\label{subsec:solvers}
A total of eight solvers were submitted to this year's competition, two of the solvers were submitted with two configurations. Table~\ref{tbl:solvers-in-tracks} summarizes which solver participated in which track and in how many configurations. Thus a total of 3 solvers participated in both the LIA and the INV tracks. In the general track participated 5 solvers, out of which 2 with two configurations. Table~\ref{tbl:solvers-in-tracks} lists all the submitted solvers and their authors.


\begin{table}[t]
\begin{center}
\begin{tabular}{r||rrrrrrrr}
 & \multicolumn{8}{c}{Solvers} \\
 Tracks & \rotatebox{90}{\alc} & \rotatebox{90}{\alccsdt} & \rotatebox{90}{\cvc} & \rotatebox{90}{\enum} & \rotatebox{90}{\ice} & \rotatebox{90}{\skac} & \rotatebox{90}{\toast} & \rotatebox{90}{\stoch} \\ \hline \hline
 LIA & 1 & 1 & 1 & 0 & 0 & 0 & 0 & 0\\
 INV & 1 & 0 & 1 & 0 & 1 & 0 & 0 & 0\\
 General &  0 & 0 & 1 & 1 & 0 & 2 & 2 & 1\\ 
\end{tabular}
\end{center}
\caption{Solvers participating in each track}
\label{tbl:solvers-in-tracks}
\end{table}

Six of the solvers are based on the CEGIS approach: the \emph{enumerative} solver (\enum), the \emph{stochastic} solver (\stoch), the \emph{Sketch} solver (\skac), the \emph{abstract solution Analyzing Synthesis Tool} (\toast), the \emph{Alchemist CS} solver (\alc), and the \emph{Alchemis CSDT} solver (\alccsdt).
 The approaches of these solvers differ in the way the candidate expression is chosen. 

The \emph{enumerative} solver enumerates the candidate solutions by increasing length, and tries the smallest candidate that was not refuted so far. In building new expressions it uses a classification to equivalence classes based on correctness on the current set of examples. For further details on the enumerative solver see~\cite{AlurBDF0JKMMRSSSSTU15,AlurBJMRSSSTU13}.

The \emph{stochastic} solver chooses the candidate  expression by a probabilistic walk on a graph where nodes are expressions and edges capture single-edits of the expression parse tree. It uses the Metropolis-Hastings Algorithm using a score function that is inversely proportional to the number of instances in the current set of examples, on which the expression is incorrect. 
For further details see~\cite{AlurBDF0JKMMRSSSSTU15,AlurBJMRSSSTU13}.

The \emph{Sketch} solver follows a symbolic approach to CEGIS and uses an SMT solvers to choose the next candidate expression~\cite{Solar-LezamaRBE05}. The \emph{Sketch Adaptive-Concretization} combines symbolic and explicit CEGIS approaches~\cite{JeonQSF15} by randomly concretizing highly influential unknowns. This approaches lends itself to parallelism. Indeed \skac\ is the only solver that used all four cores in this year's competition (all the other solvers used a single core).

The \toast\ solver refines the enumerative CEGIS approach by considering \emph{abstract expressions}; these are expressions that combine grammar non-terminals and concrete symbols. The idea is to use an SMT solver to refute impossible abstract expressions, instead of refuting all their respective concretizations. A check of the feasbility of an abstract expression to be a solution involves qunatifer alternation. The SMT calls are thus bounded by a small time unit (10 seconds)~\cite{SosyToast}.

The tools \alc~\cite{AlchemistCS} and \alccsdt~\cite{AlchemistCSDT} are specialized solvers for the LIA track. They are base on the Alchemist~\cite{Saha0M15}, they both synthesize nested \emph{if-then-else} expressions, 
and combine enumeration and constraint-solver based technique to find leaf expressions
The \alccsdt\ works in two modes, in the first all variables (qunatified and unquantified) are assumed to be integers, and in the second variables may be either integers or Booleans. 
The \alc\ tool can only solve \emph{point-wise} SyGuS problems. A SyGuS problem is said to be point-wise if the following holds: let $\mathcal{F}$ be the class of all functions that are solutions to the specification; then any function $h$ that maps each input to an output consistent to some function in $\mathcal{F}$, is also a solution to the specification.

\begin{table}[t]
\begin{center}
\begin{tabular}{r||l}
 Solver &  Authors \\ \hline \hline
 \alc & Daniel Neider (UIUC),
        Shambwaditya Saha (UIUC) and
        P. Madhusudan (UIUC) \\
 \alccsdt & Shambwaditya Saha (UIUC),
          Daniel Neider (UIUC) and
          P. Madhusudan (UIUC) \\
 \cvc & Andrew Reynolds (EPFL),
      Viktor Kuncak (EPFL), 
       Cesare Tinelli (Univ. of Iowa), \\ 
       & Clark Barrett (NYU), 
       Morgan Deters (NYU) and
       Tim King (Verimag) \\
 \enum & Abhishek Udupa (Penn) \\
 \ice & Daniel Neider (UIUC), 
       P. Madhusudan (UIUC) and
       Pranav Garg (UIUC) \\
 \skac  & Jinseong Jeon (UMD),	Xiaokang Qiu (MIT),	Armando Solar-Lezama (MIT) and\\
       & 	Jeffrey S. Foster (UMD)	\\
 \toast  & Heinz Riener (DLR) and
          Ruediger Ehlers (DLR) \\
 \stoch & Mukund Raghothama (Penn) \\              
\end{tabular}
\end{center}
\caption{Submitted solvers and their authors}
\label{tbl:solvers-authors}
\end{table}

The \ice~\cite{ICEDT} tool is a specialized solver for the invariant synthesis track. It builds on the ICE learning approach~\cite{GargLMN14} --- an invariant synthesis approach in which the teacher responses to a given hypothesis may be concrete examples, counterexamples or implication.  \ice\ extends the ICE approach by using decision trees as suggested in~\cite{GargNMR15}. 

The \cvc\ solver is the only solver that is implemented inside an SMT solver~\cite{ReynoldsDKTB15}. It reduces the synthesis problem to an unsatisfibility of quantified SMT formulae. Indeed, as noted in~\cite{AlurBJMRSSSTU13,AlurBDF0JKMMRSSSSTU15}, for linear integer arithmetic problems,  suppose the synthesis problem involves two variables $x$ and $y$, then it can be articulated as the following quantified SMT satisfiability query: 
$\exists a, b, c.~ \forall x, y. ~ \varphi[f/ax+by+c]$.
Since traditional SMT solvers are focused on instnatiation-based methods to show \emph{unsatisfiability}, the \cvc\ tool reformulates LIA problems as unsatisfiability of the synthesis conjecture $\forall f, \exists x_0,x_1,\ldots,x_i,\neg\varphi[f,x_0,x_1,\ldots,x_i]$. For non-LIA problems, the \cvc\ uses an encoding  of the syntax restrictions using first order variables and uses a deep embedding into an extension of the background theory $T$ with a theory of algebraic data types.

\subsection{Experimental Setup}
\label{subsec:tech}
The solvers were run on the StarExec platform~\cite{starexec} with a dedicated cluster of 12 nodes, where each node consisted of two 4-core 2.4GHz Intel processors with 256GB RAM and a 1TB hard drive. The memory usage limit of each solver run was set to 128GB. The wallclock time unit was set to 3600 seconds (thus, a solver that used all cores could consume at most 14400 seconds cpu time).
 
The solutions that the solvers produce are being checked for both syntactic and semantic correctness. That is, a first post-processor checks that the produced expression adheres to the grammar specified in the given benchmark, and if this check passes, a second post-processor checks that the solution adheres to semantic constraints given in the benchmark (by invoking an SMT solver).

\section{Competition Results and Analysis}
\label{sec:results}
In this section we present the  results of SyGuS-COMP'15. We start with an overview of the results in subsection~\ref{subsec:bench-res-overview}. 
In subsection~\ref{subsec:bench-features} we try to understand the different features a certain benchmark may or may not have, and analyze each benchmark category in terms of the mentioned features. In subsection~\ref{subsec:benchs-pres} we provide detailed results of the competition --- going category by category, we present for each benchmark the range of times the solver took to solve that benchmark and which solver was the fastest, and the range of obtained expression sizes, and which solver produced the smallest expression. Finally, we conclude in subsection~\ref{subsec:obs} with some interesting observations and conjectures.

\subsection{Results Overview} 
\label{subsec:bench-res-overview}
\begin{figure*}
\noindent\makebox[\textwidth]{
\scalebox{0.6}{
\begin{tabular}{c}
\includegraphics[width=8in]{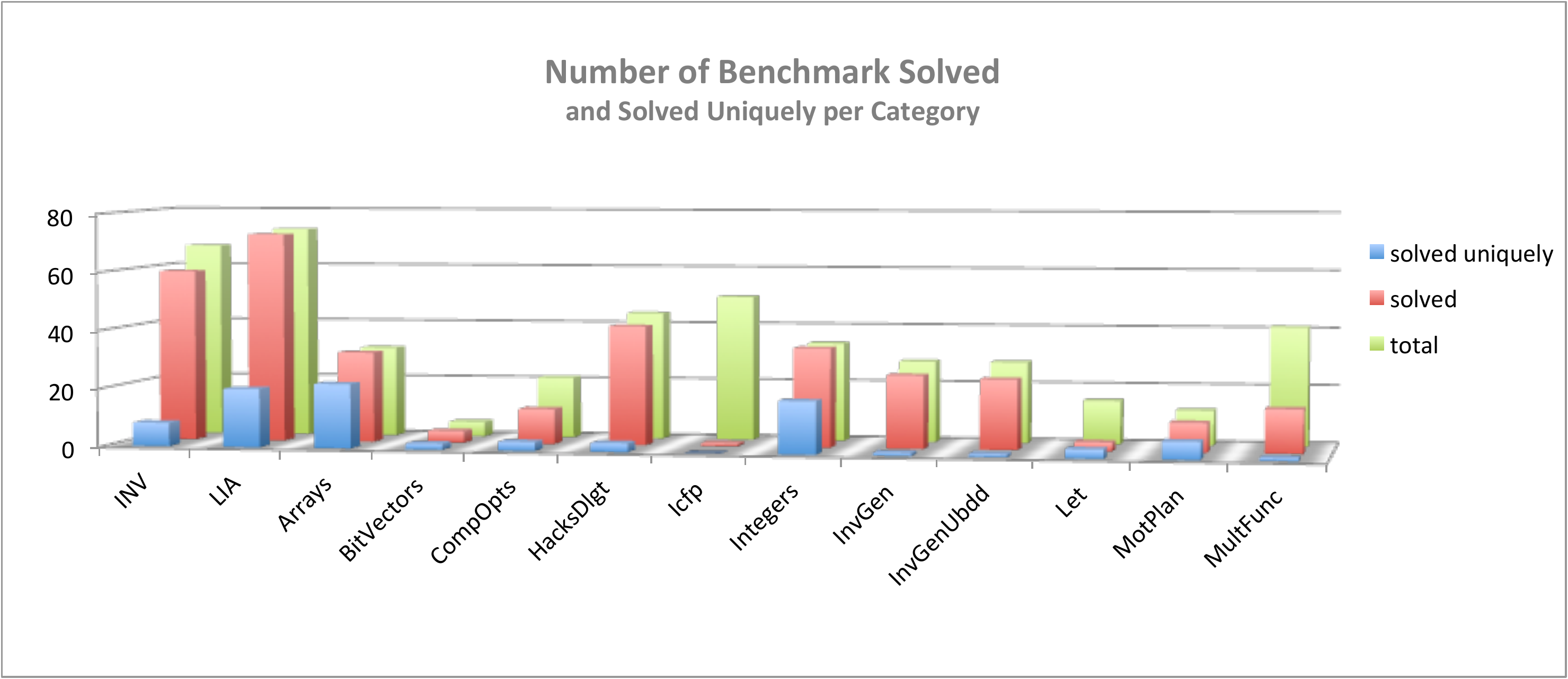} 
\end{tabular}
}}
\caption{Results on the number of solved benchmarks.}\label{fig:total-results}
\end{figure*}


Figure~\ref{fig:total-results} shows for each track and category (a) the total number of benchmarks (in green), (b) the number of benchmarks solved by at least one solver (in red), and (c) the number of benchmarks solved by only one of the solvers (in blue).

\begin{figure}
\noindent\makebox[\textwidth]{
\scalebox{0.6}{
\begin{tabular}{c}
\includegraphics[width=8in]{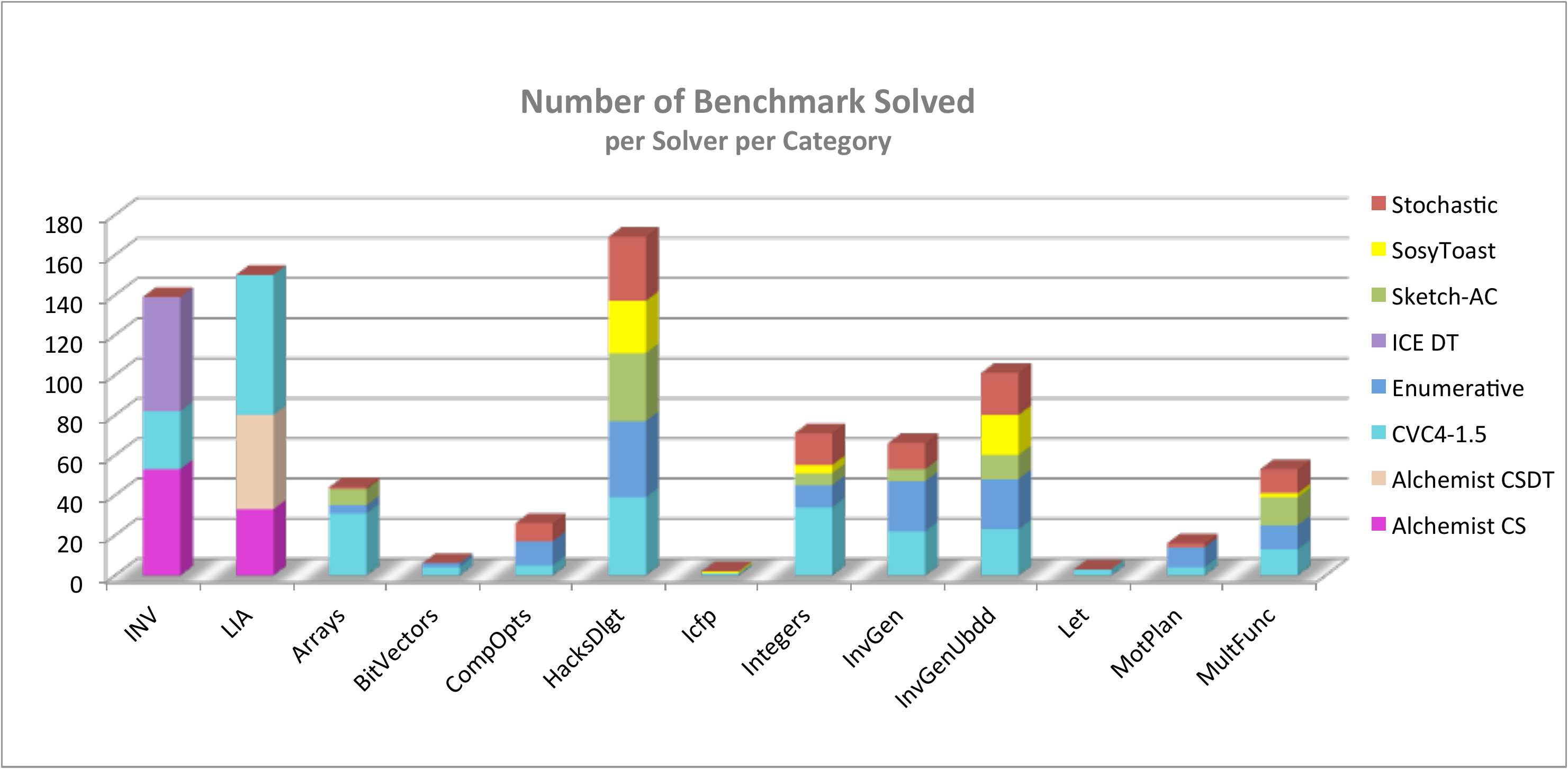} \\
\includegraphics[width=8in]{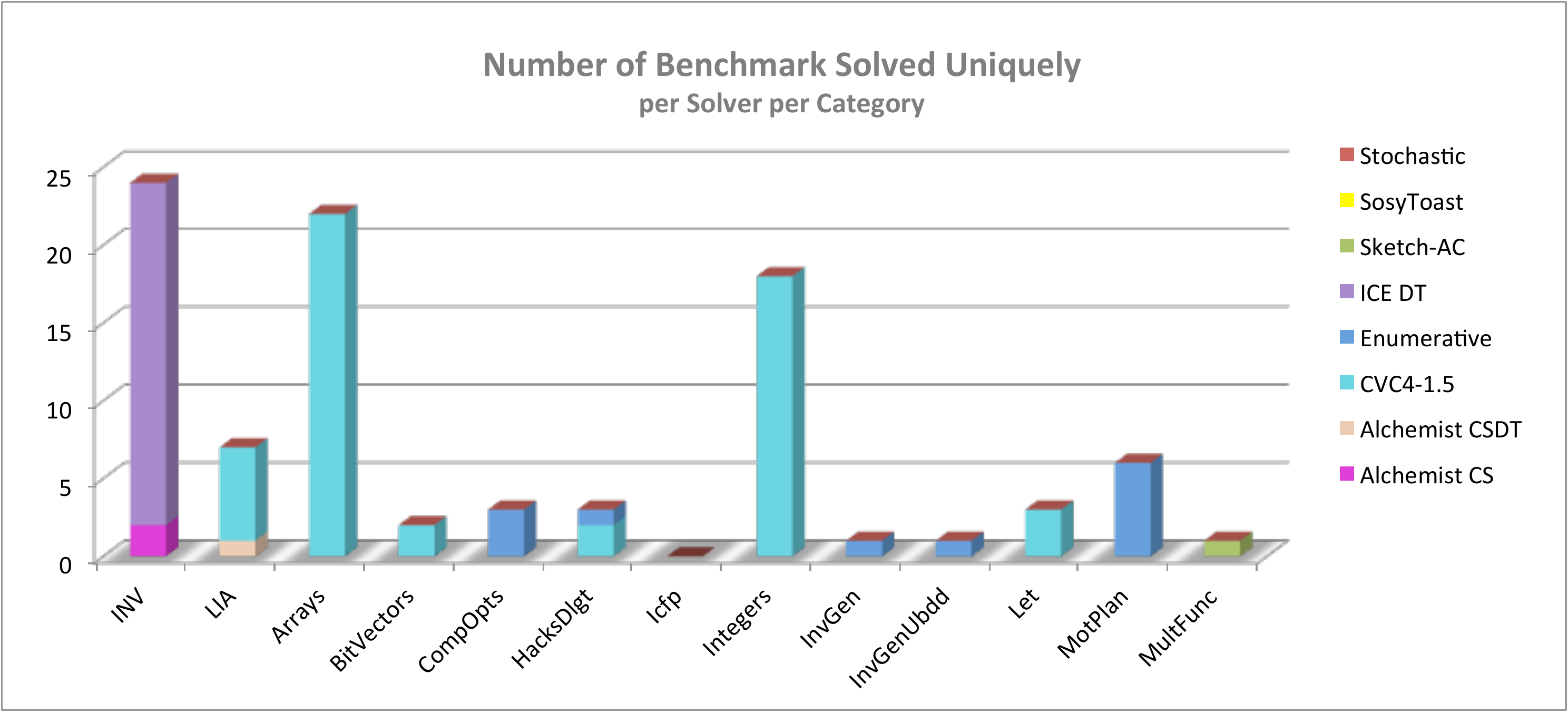} 
\end{tabular}
}}
\caption{Results on the number of solved benchmarks per solver per category.}\label{fig:solver-results}
\end{figure}

\label{subsec:solvers-res}
Figure~\ref{fig:solver-results} on the top shows for each track and category the number of benchmarks solved by each of the solvers. It is evident from the figure that on the INV track, \ice\ solved more instances than the others; it solved 57 out of 67 benchmarks. The Alchemist solver solved 53 benchmarks in the INV track. 

In the LIA track the leading solver is \cvc, which solved 70 out of 73 benchmarks. The second place goes to \alccsdt\ which solved 47 instances.

\cvc\ is also the winner in the general track, in which it solved 179 out of 309 benchmarks. The second place in this track goes to the \enum\ solver, the winner of last year's competition, which solved 139 instances. The third place goes to the \stoch\ solver, which solved 109 instances.

Figure~\ref{fig:solver-results} on the bottom shows for each track and category the number of benchmarks solved uniquely by each of the solvers. We say that a solver solved a benchmark uniquely if it is the only solver to solve that benchmark. It is evident that on the INV track, the \ice\ solver solved more benchmarks uniquely, whereas in the LIA track the \cvc\ solver solved more benchmarks uniquely. In the general track, both \cvc\ and \enum\ solved many benchmarks that all the other solvers did not. It is interesting to see that for each benchmark category, there is a clear winner in terms of the number of benchmarks solved uniquely.

\subsection{Benchmarks' Features} 
\label{subsec:bench-features}
Since the solving strategies for synthesizers can be greatly influenced by the problem representation, we quantify here some different features that are exhibited by the benchmarks from different categories in Table~\ref{tbl:features}.\footnote{The mentioned features may not exist in \emph{all} benchmarks of the category, but they are present in most of the benchmarks in that category.} We consider here four main categories: 1) whether the invocation of the unknown function is always present with the same set of arguments (\emph{single invocation}) or there are multiple different function invocations (\emph{multiple invocation}), 2) whether there is a single unknown function (\emph{single unknown}) or multiple unknown functions (\emph{multiple unknown}), 3) whether the specification constraint is complete (\emph{complete spec}) or partially specified using input-output examples (\emph{partial spec}), and finally 4) whether the grammar for the unknown function is \emph{restricted}, \emph{semi-general}, or \emph{general}.

\paragraph{Function Invocation} We say a benchmark belongs to the \emph{single invocation} class if the unknown function is called (possibly multiple times) with the same set of arguments in the specification constraint. For example, consider the specification constraint for the \texttt{max4} benchmark. This benchmark belongs to the single invocation class because the \texttt{max4} function is always called with the same set of arguments \texttt{x}, \texttt{y}, \texttt{z}, and \texttt{w}.

\begin{verbatim}
(constraint (>= (max4 x y z w) x))
(constraint (>= (max4 x y z w) y))
(constraint (>= (max4 x y z w) z))
(constraint (>= (max4 x y z w) w))
(constraint (or (= x (max4 x y z w))
            (or (= y (max4 x y z w))
            (or (= z (max4 x y z w))
                (= w (max4 x y z w))))))

\end{verbatim}

Similarly, we say a benchmark belongs to the \emph{multiple invocation} class if the unknown function is called multiple times with different set of arguments. For example, consider the \texttt{icfp\_7\_10} benchmark, where the unknown function \texttt{f} is called multiple times with different set of arguments.

\begin{verbatim}
(constraint (= (f #x1be88589ba201842) #xe4177a7645dfe7bd))
(constraint (= (f #x49ea2ae53e599623) #x93d455ca7cb32c46))
(constraint (= (f #xea82cc5e6104247d) #xea82cc5e6104247d))
(constraint (= (f #x75820d31bed79b87) #xeb041a637daf370e))
(constraint (= (f #xe682665199ee31a8) #x197d99ae6611ce57))
\end{verbatim}

\paragraph{Number of Unknown Functions} We say a benchmark belongs to the class of \emph{single unknown} class if there is only one unkown function that needs to be synthesized. For example, the \texttt{max4} and \texttt{icfp\_7\_10} benchmarks above require to synthesize a single function \texttt{max4} and \texttt{f} respectively. The benchmarks that require to synthesize multiple unknown functions are said to belong to the \emph{multiple unkown} class. For example, consider the specification constraint from the \texttt{s8} benchmark that requires to synthesize three unknown functions \texttt{f1}, \texttt{f2}, and \texttt{f3}, such that their sum is equal to a given linear arithmetic expression.

\begin{verbatim}
(constraint (= (+ (+ (f1 x y z) (f2 x y z)) (f3 x y z)) (+ (+ x y) z)))
(constraint (= (f2 x y z) (- y 1)))
\end{verbatim}

We note that multiple unknown functions can be theoretically represented as a single unknown function by taking the union of respective grammars of the functions and adding a new \texttt{Start} non-terminal. But, we still distinguish benchmarks into different categories based on number of unknown functions as some solvers might exploit the additional structural information present in the benchmarks.

\begin{table*}[]
\centering
\label{tbl:features}
\begin{tabular}{lllll}
\hline
\hline
{\bf Benchmark}   & {\bf Function} & {\bf \# Unkown} & {\bf Specification} & {\bf Grammar} \\
{\bf Category} & {\bf Invocations} & {\bf Functions} & {\bf Constraint} & {\bf Generality} \\
\hline
\hline
{\bf Array}                & Single              & Single              & Complete                 & Restricted         \\ \hline
{\bf Let}                  & Single              & Multiple            & Complete                 & Restricted         \\ \hline
{\bf Invariant Generation} & Multiple            & Single              & Complete                 & General            \\ \hline
{\bf ICFP}                 & Multiple            & Single              & Partial                  & General            \\ \hline
{\bf HackerDel-d0}         & Single              & Single              & Complete                 & Restricted         \\ \hline
{\bf HackerDel-d1}         & Single              & Single              & Complete                 & Semi-General       \\ \hline
{\bf HackerDel-d5}         & Single              & Single              & Complete                 & General            \\ \hline
{\bf Integer}              & Single              & Single              & Complete                 & Semi-General       \\ \hline
{\bf Bitvector}            & Single              & Single              & Complete                 & Semi-General       \\ \hline
{\bf CompilerOpts}             & Single              & Single              & Complete                 & Restricted         \\ \hline
{\bf MultipleFuncs}        & Multiple            & Multiple            & Partial                  & Restricted         \\ \hline
{\bf Motion Planning}      & Multiple            & Single              & Complete                 & Semi-General       \\ \hline
{\bf LIA Track}                  & Single              & Single              & Complete                 & General            \\ \hline
{\bf INV Track}                  & Multiple            & Single              & Partial                  & General           \\  \hline
\end{tabular}
\caption{The list of features present in benchmarks from different categories.}
\end{table*}

\paragraph{Specification Constraint} Since the specification language for the SyGuS-IF is based on constraints and is therefore quite general, the specification for different benchmarks can be defined with varying levels of completeness. We say a benchmark belongs to the \emph{partial spec} class if the specification constraint allows for multiple possible \emph{semantic} solutions for the unkown functions. We say two functions are semantically different if there exists some input for which they produce different outputs.  For example, the specification for the \texttt{icfp\_7\_10} benchmark is partial as it is specified using a set of input-output examples. The specification for the \texttt{s8} benchmark shown above is also partial as it allows for multiple possible functions \texttt{f1}, \texttt{f2}, and \texttt{f3}.

On the other hand, we say a benchmark belongs to the \emph{complete spec} class if the complete functional specification of the unknown function is provided. For example, consider the specification for the compiler optimization benchmark \texttt{qm\_loop\_1}, where the constraint on the unknown function \texttt{qm-loop} fully specifies the result of the function for every input.

\begin{verbatim}
(set-logic LIA)

(define-fun qm ((a Int) (b Int)) Int (ite (< a 0) b a))

(synth-fun qm-loop ((x Int)) Int
    ((Start Int (x 0 1 3
                 (- Start Start)
                 (+ Start Start)
                 (qm Start Start)))))

(declare-var x Int)

(constraint (= (qm-loop x) (ite (<= x 0) 3 (- x 1)))) 

(check-synth)
\end{verbatim}

\paragraph{Grammar Generality} We divide the benchmarks into three categories based on the generality of the grammars for the unknown functions. The \emph{restricted grammar} class consists of benchmarks where the grammars are restricted to only contain operators and variables that are strictly necessary for the unknown function. The second class of grammar, \emph{semi-general grammar}, consists of a few more choices for operators and rules that might not be necessarily needed for synthesizing the unknown function. The third class of grammar, \emph{general grammar}, consists of a very general grammar that consists of all possible operators and variables from a given theory. For example, consider the grammars for the \texttt{hd-17-d0-prog} (restricted grammar), \texttt{hd-17-d1-prog} (semi-general), and \texttt{hd-17-d5-prog} (general grammar) benchmarks where all the constraints are same except for the generality of the grammar. The benchmark \texttt{hd-17-prog} is taken from the Hacker's Delight book that needs to synthesize a bitvector function to turn-off the rightmost contiguous string of 1 bits in the input bitvector \texttt{x}.

\begin{verbatim}
;; hd-17-d0-prog

(synth-fun f ((x (BitVec 32))) (BitVec 32)
    ((Start (BitVec 32) ((bvand Start Start)
                         (bvadd Start Start)
                         (bvsub Start Start)
                         (bvor Start Start)
                          x
                         #x00000001))))


\end{verbatim}

\begin{verbatim}
;; hd-17-d1-prog

(synth-fun f ((x (BitVec 32))) (BitVec 32)
    ((Start (BitVec 32) ((bvand Start Start)
                         (bvadd Start Start)
                         (bvxor Start Start)
                         (bvsub Start Start)
                         (bvor Start Start)
                         (bvnot Start)
                         (bvneg Start)
                         x
                         #x00000001
                         #x00000000
                         #xFFFFFFFF))))

\end{verbatim}

\begin{verbatim}
;; hd-17-d5-prog

(synth-fun f ((x (BitVec 32))) (BitVec 32)
    ((Start (BitVec 32) ((bvnot Start)
                         (bvxor Start Start)
                         (bvand Start Start)
                         (bvor Start Start)
                         (bvneg Start)
                         (bvadd Start Start)
                         (bvmul Start Start)
                         (bvudiv Start Start)
                         (bvurem Start Start)
                         (bvlshr Start Start)
                         (bvashr Start Start)
                         (bvshl Start Start)
                         (bvsdiv Start Start)
                         (bvsrem Start Start)
                         (bvsub Start Start)
                         x
                         #x0000001F
                         #x00000001
                         #x00000000
                         #xFFFFFFFF))))


\end{verbatim}

The grammars for the unknown functions in the Conditional Linear Arithmetic (LIA) track and the Invariant Synthesis (INV) track belong to the class of general grammars, where the following grammar is used for all unknown functions in the benchmarks.

\begin{verbatim}
(synth-fun f ((x1 Int) ... (xn Int)) Int
    ((Start Int (StartInt))
     (StartInt Int (x1 ... xn ConstantInt
        (+ StartInt StartInt)
        (- StartInt StartInt)
        (* StartInt ConstantInt)
        (* ConstantInt StartInt)
        (div StartInt ConstantInt)
        (mod StartInt ConstantInt)
        (ite StartBool StartInt StartInt)))
     (ConstantInt (Constant Int))
     (StartBool Bool (true false
        (and StartBool StartBool)
        (or StartBool StartBool)
        (=> StartBool StartBool)
        (xor StartBool StartBool)
        (xnor StartBool StartBool)
        (nand StartBool StartBool)
        (nor StartBool StartBool)
        (iff StartBool StartBool)
        (not StartBool)
        (= StartBool StartBool)
        (<= StartInt StartInt)
        (= StartInt StartInt)
        (>= StartInt StartInt)
        (> StartInt StartInt)
        (< StartInt StartInt)))))
\end{verbatim}

\subsection{Detailed Results}
\label{subsec:benchs-pres}
In the following section we show the results of the competition from the benchmark's perspective. 
For a given benchmark we would like to know: how many solvers solved it, what is the min and max time to solve,  what are the min and max size of the expressions produced, which solver solved the benchmark the fastest, and which solver produced the smallest expression.

\begin{figure*}
\noindent\makebox[\textwidth]{
\scalebox{0.6}{
\begin{tabular}{c}
\includegraphics[width=9.5in]{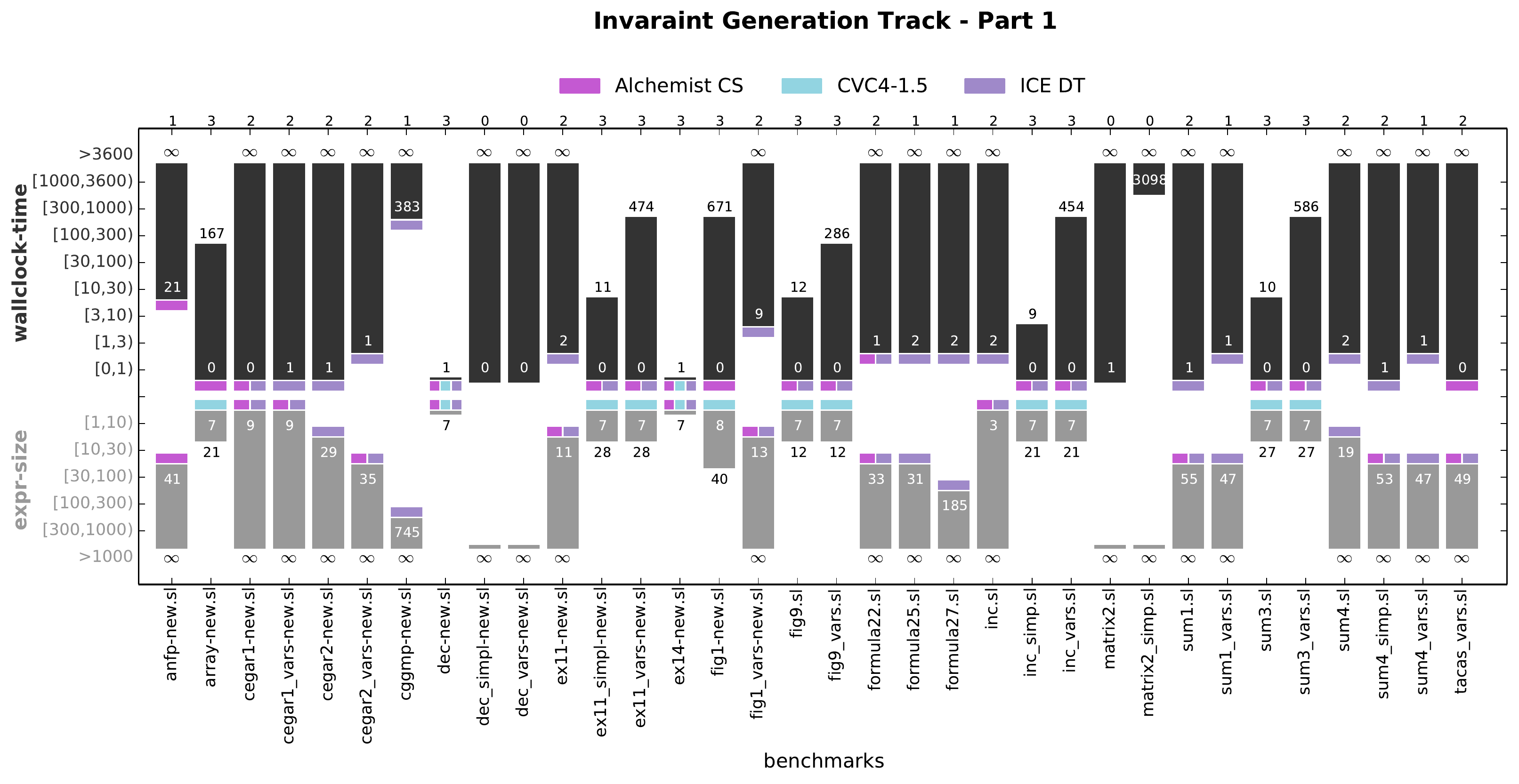} \\
\includegraphics[width=9.5in]{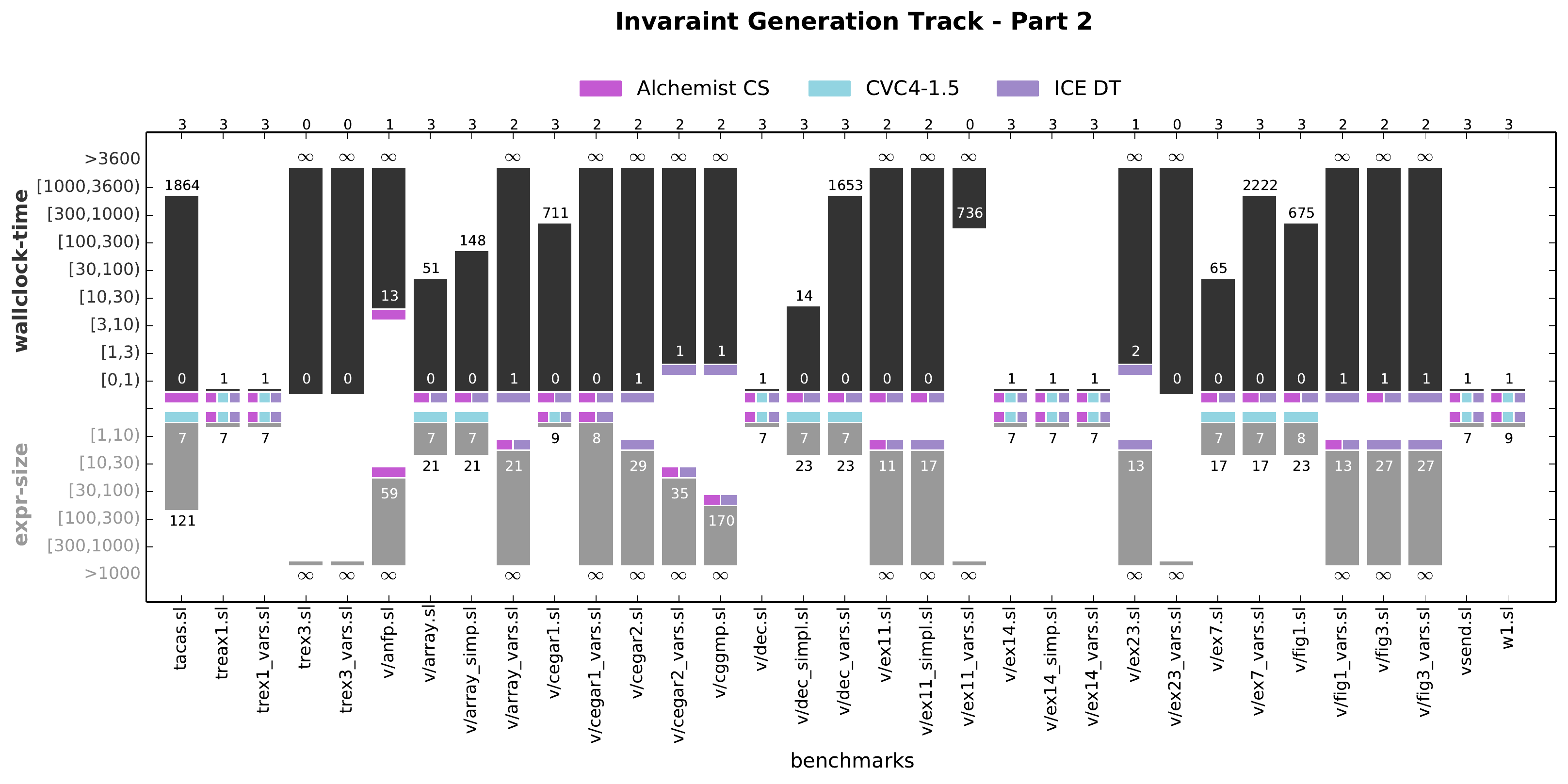} 
\end{tabular}
}}
\caption{Evaluation of INV benchmarks.}\label{fig:inv-results}
\end{figure*}

We represents the results per benchmarks in groups organized per tracks and categories. For instance, Fig.~\ref{fig:inv-results} shows all benchmarks of the INV tracks (first half in the upper figure, and second half in the lower figure). The black bars show the range of the time to solve among the different solvers in psuedo logarithmic scale (as indicated on the upper part of the y-axis). Inspect for instance benchmark \texttt{anpf-new.sl}. The black bar indicates that the fastest solver to solve it used between 10 to 30 second, and the slowest did not solve it within the given time bound (3600 second). 
The black number above the black bar indicates the exact number of seconds (floor-rounded to the nearest second) it took the slowest solver to solve a benchmark (and $\infty$ if at least one solver exceeded the time bound). Thus, we can see that the slowest solver to solve \texttt{array-new.sl} took 167 seconds to solve it. The white number at the lower part of the bar indicates the time of the fastest solver to solve that benchmark. Thus, we can see that the fastest solver to solve \texttt{anpf-new.sl} required 21 second to do so, and the fastest solver to solve \texttt{array-new.sl} took less than a second. The colored squares/rectangles next to the lower part of the black bar, indicate which solvers were the fastest to solve that benchmark (according to the solvers' legend at the top). Here, \emph{fastest} means in the same logarithmic scale as the absolute fastest solver. For instance, we can see that \alc\ was the fastest to solve \texttt{anpf-new.sl},
that both \alc\ and \ice\ solved \texttt{cegar-new.sl} in less than a second, and that all solvers (\ice,\alc, and \cvc) solved \texttt{dec-new.sl} in less than a second. 

Similarly, the gray bars indicate the range of expression sizes in psuedo logarithmic scales (as indicated on the lower part of the y-axis), where the size of an expression is determined by the number of nodes in its parse tree.
The black number at the bottom of the gray bar indicates the exact size expression of the largest solution (or $\infty$ if it exceeded 1000), and the white number at the top of the gray bar indicates the exact size expression of the smallest solution. The colored squares/rectangles next to the upper part of the gray bar indicates which solvers (according to the legend) produced the smallest expression (where \emph{smallest} means in the same logarithmic scale as the absolute smallest expression). For instance, for \texttt{tacas.sl} the smallest expression produced had size 7, the biggest expression had size 121, and the solver which produced the smallest expression is \cvc.  For \texttt{treax1.sl} all solvers produced an expression of size smaller than 10, and the absolute value of the smallest expression was 7.

Finally, at the top of the figure above each benchmark there is a number indicating the number of solvers that solved that benchmark. For instance, one solver solved \texttt{anpf-new.sl}, two solvers solved \texttt{cegar-new.sl}, three solvers solved \texttt{array-new.sl}, and no solver solved \texttt{matrix2.sl} or \texttt{matrix2-simp.sl}. Note that the reason \texttt{matrix2.sl} has 1 as the lower time bound, is that that is the time to terminate rather than the time to solve. Thus, one of the solvers has terminated within less than 1 second, but either it did not produce a result, or it produced an incorrect result. When no solver produced a correct result, there will be no colored squares/rectangles next to the lower parts of the bars.

\begin{figure*}
\noindent\makebox[\textwidth]{
\scalebox{0.6}{
\begin{tabular}{c}
\includegraphics[width=9.5in]{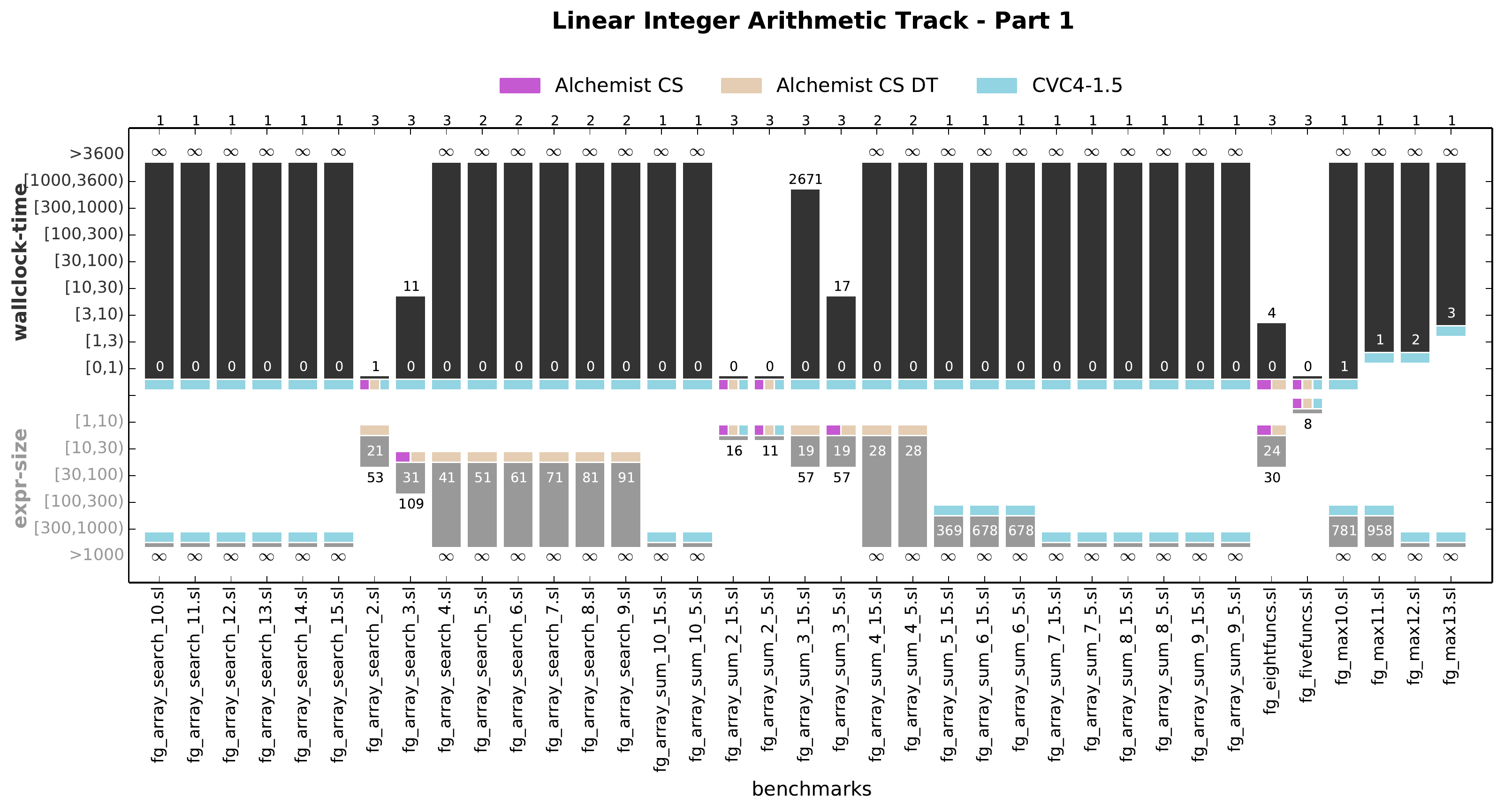} \\
\includegraphics[width=9.5in]{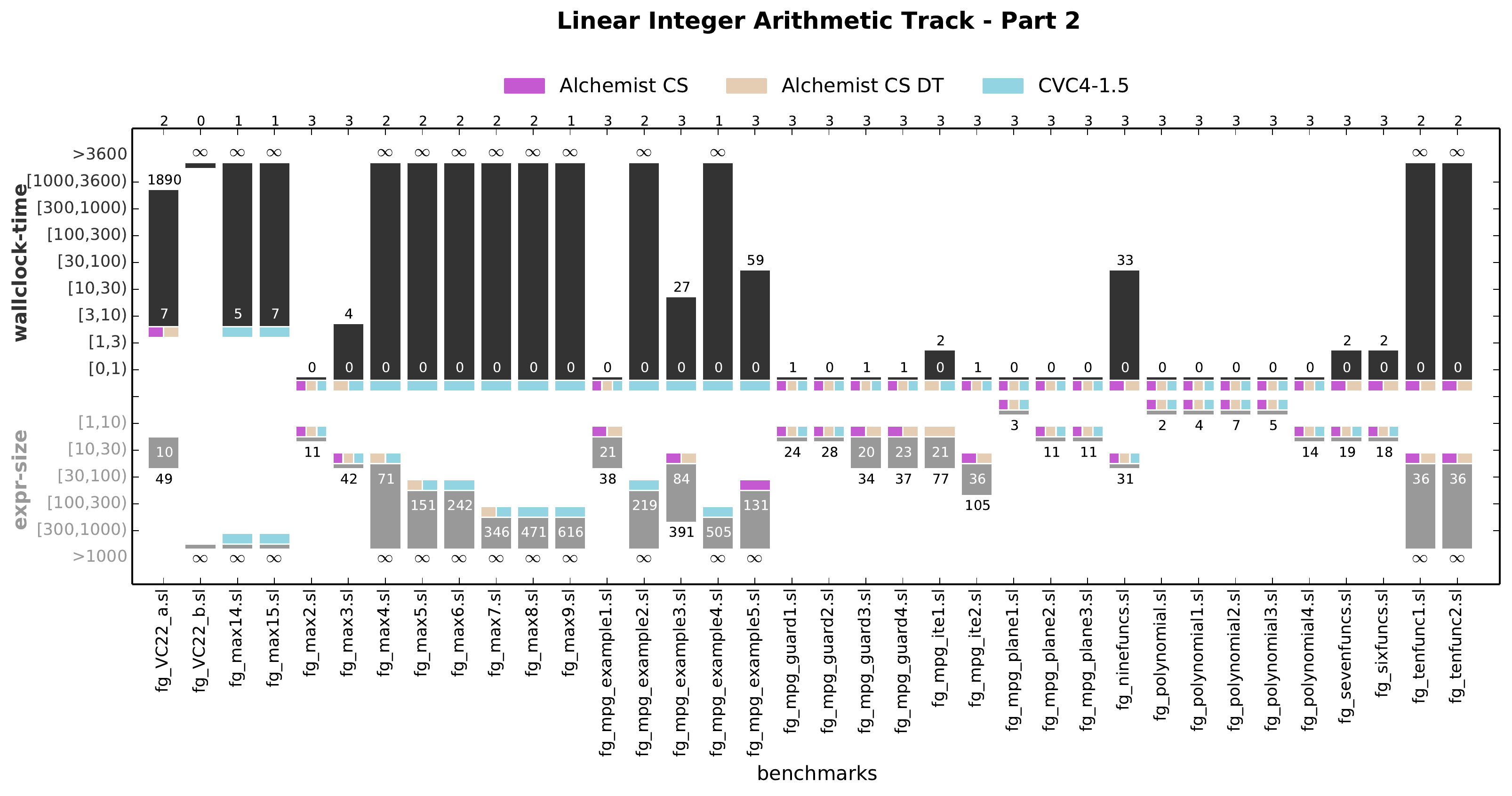} 
\end{tabular}
}}
\caption{Evaluation of LIA benchmarks.}\label{fig:lia-results}
\end{figure*}

\begin{figure*}
\noindent\makebox[\textwidth]{
\scalebox{0.6}{
\begin{tabular}{c}
\includegraphics[width=9.5in]{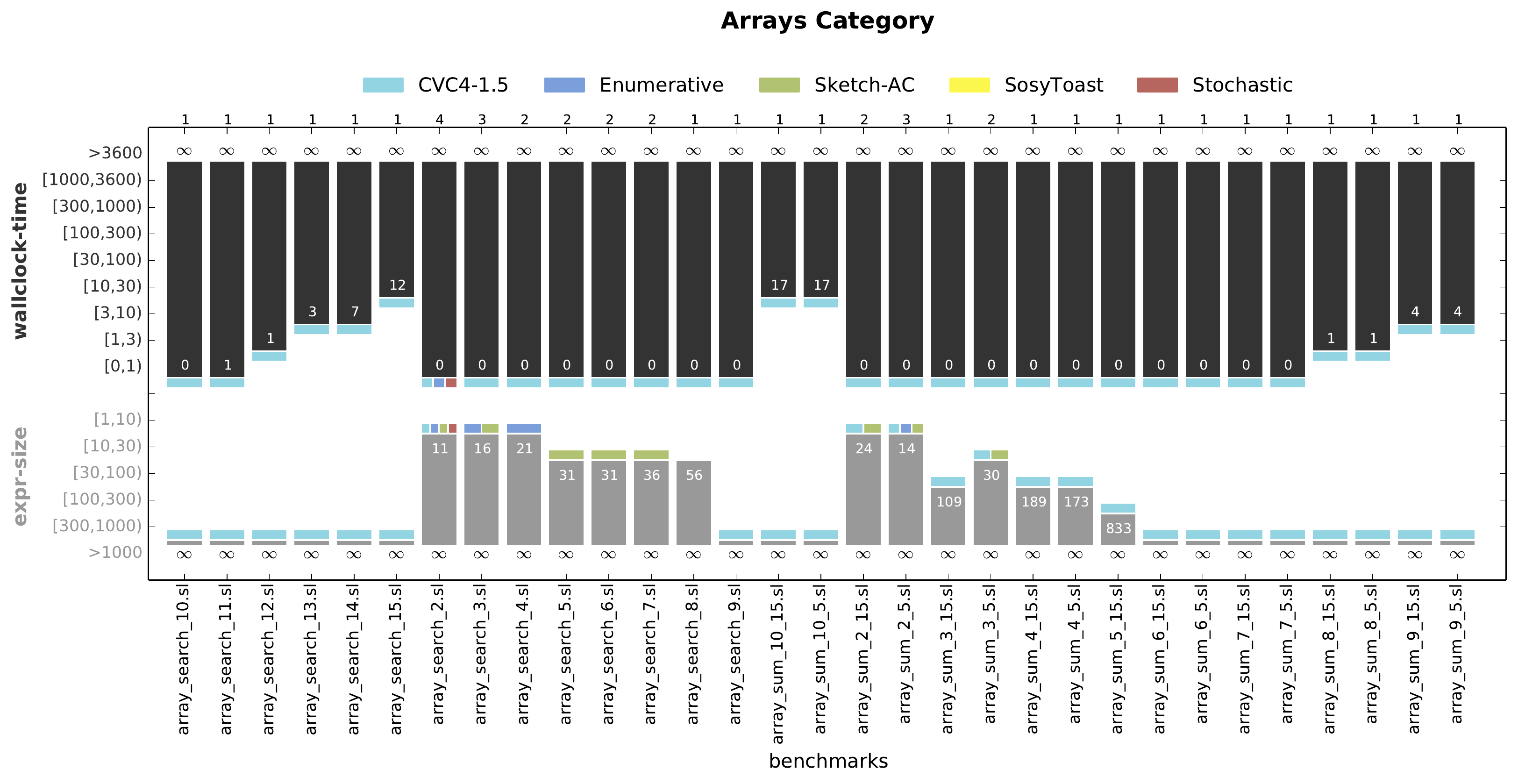} 
\end{tabular}
}}
\caption{Evaluation of Arrays benchmarks.}\label{fig:arrays-results}
\end{figure*}

\begin{figure*}
\noindent\makebox[\textwidth]{
\scalebox{0.6}{
\begin{tabular}{c}
\includegraphics[width=9.5in]{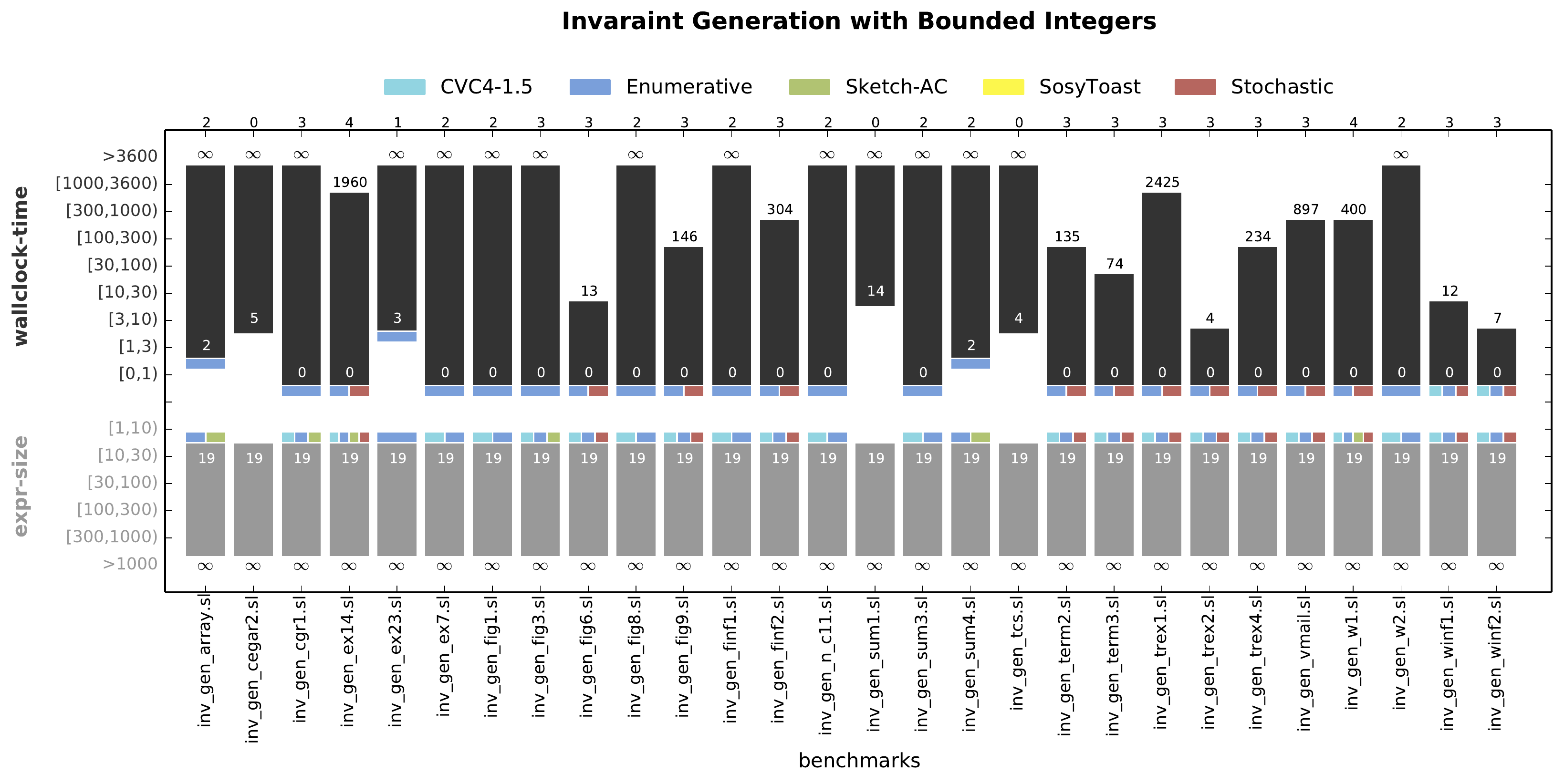} \\
\includegraphics[width=9.5in]{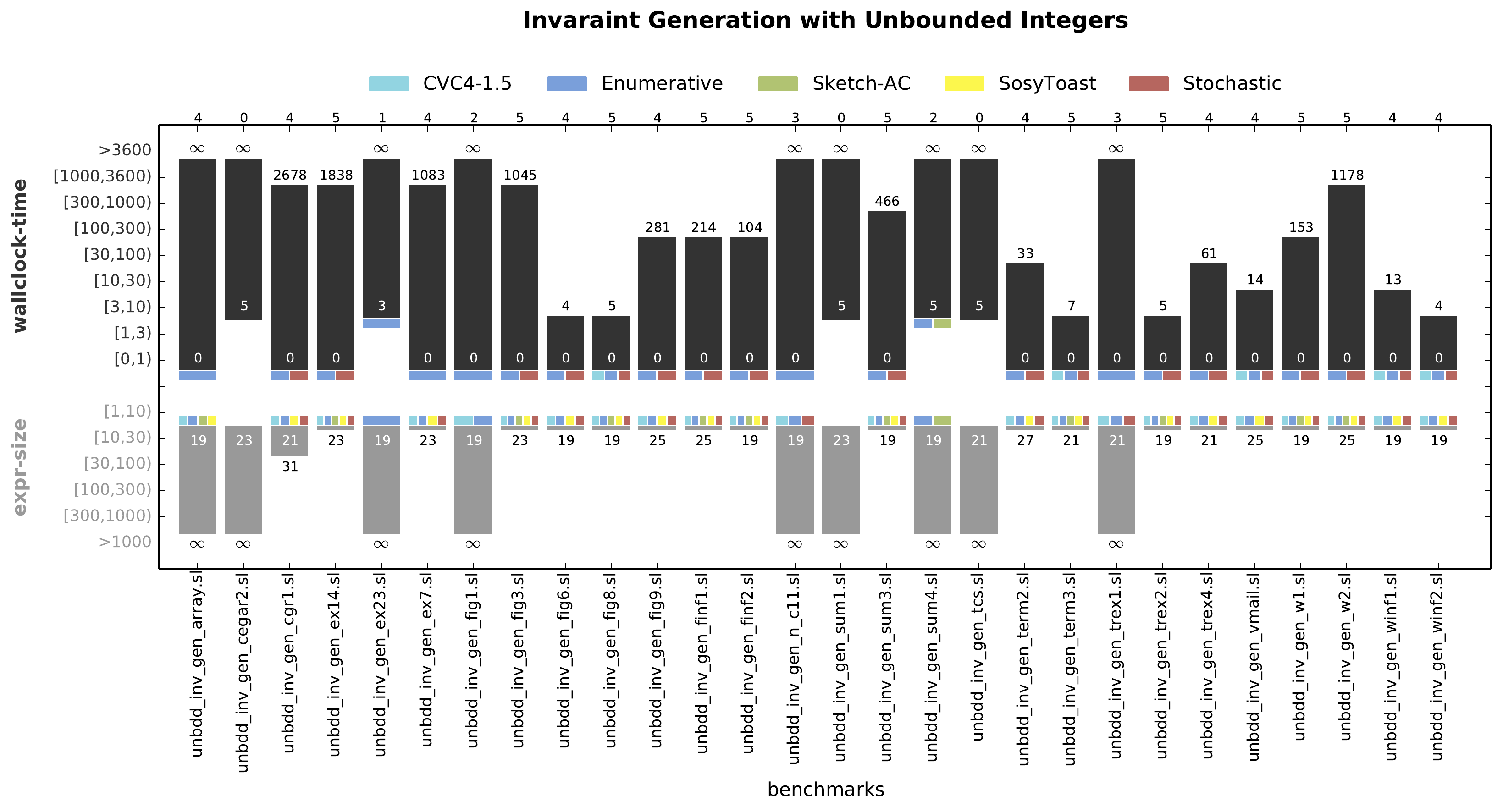} 
\end{tabular}
}}
\caption{Evaluation of Invariant Generations with/without unbounded integers benchmarks. (These are benchmarks from the general track.) }\label{fig:invgen-results}
\end{figure*}

\begin{figure*}
\noindent\makebox[\textwidth]{
\scalebox{0.6}{
\begin{tabular}{c}
\includegraphics[width=9.5in]{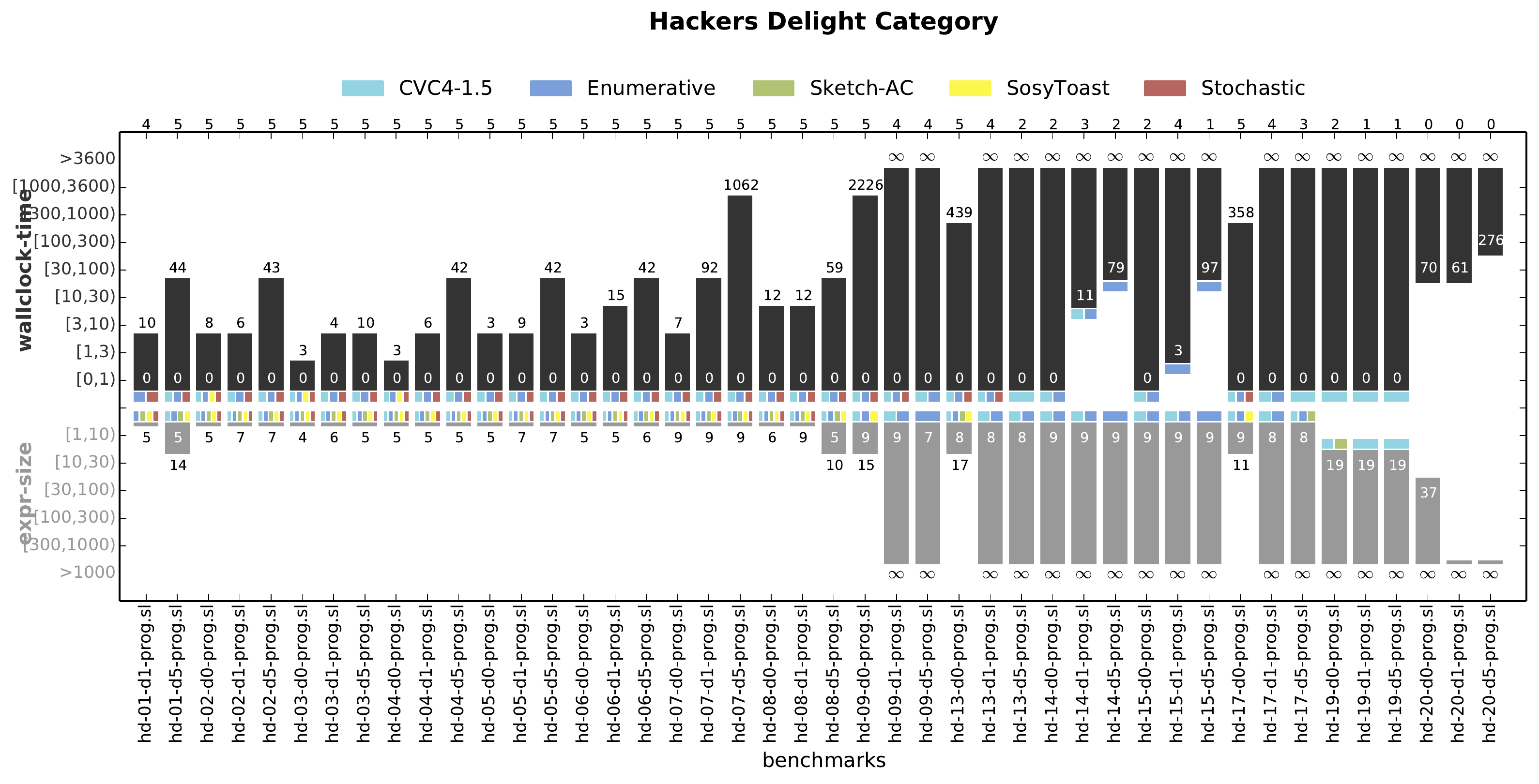} \\
\end{tabular}
}}
\caption{Evaluation of Hacker's Delight benchmarks.}\label{fig:hck-dlgt-results}
\end{figure*}

\begin{figure*}
\noindent\makebox[\textwidth]{
\scalebox{0.6}{
\begin{tabular}{c}
\includegraphics[width=9.5in]{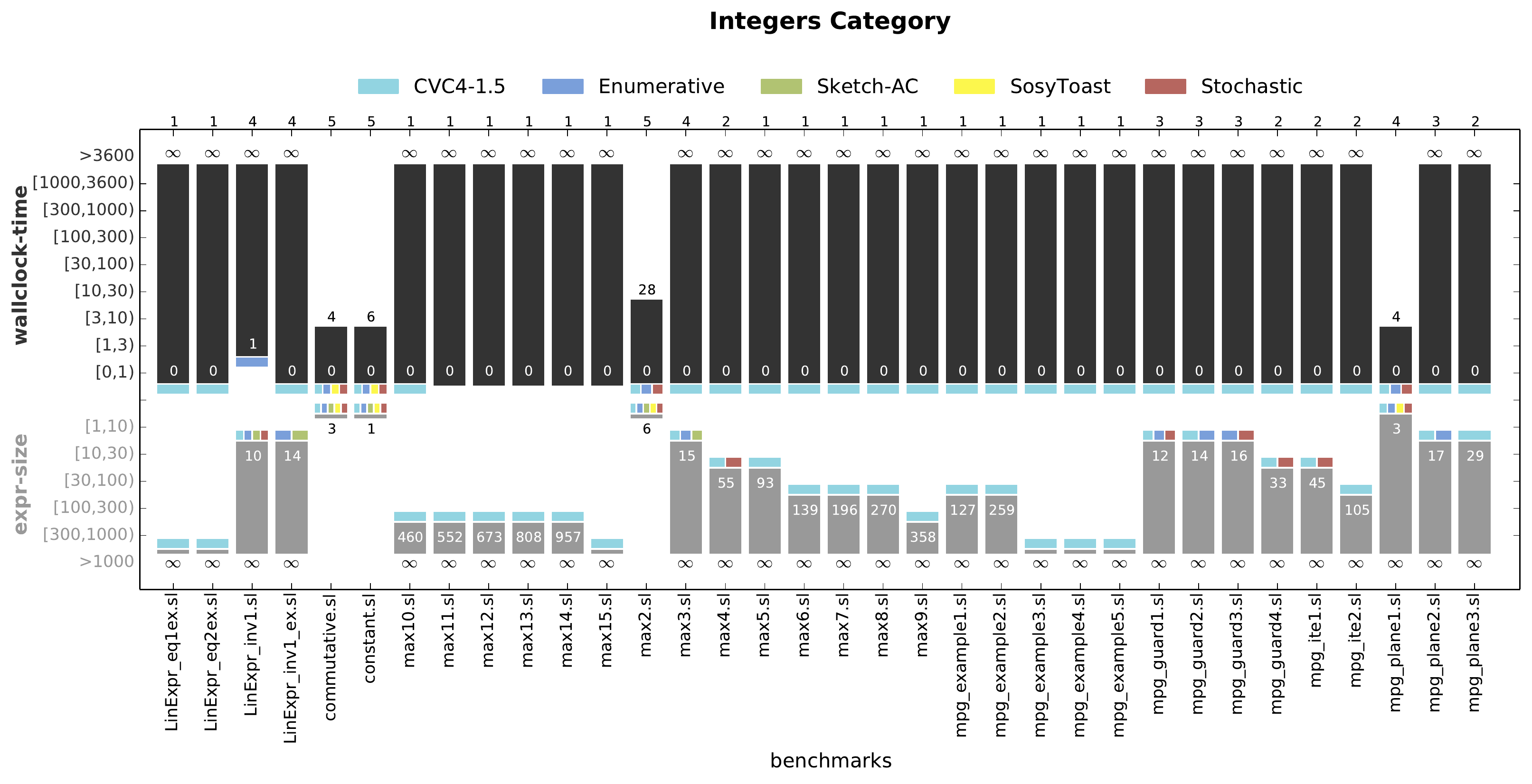} \\
\end{tabular}
}}
\caption{Evaluation of Integers benchmarks.}\label{fig:ints-results}
\end{figure*}

\begin{figure*}
\noindent\makebox[\textwidth]{
\scalebox{0.6}{
\begin{tabular}{c}
\includegraphics[width=9.5in]{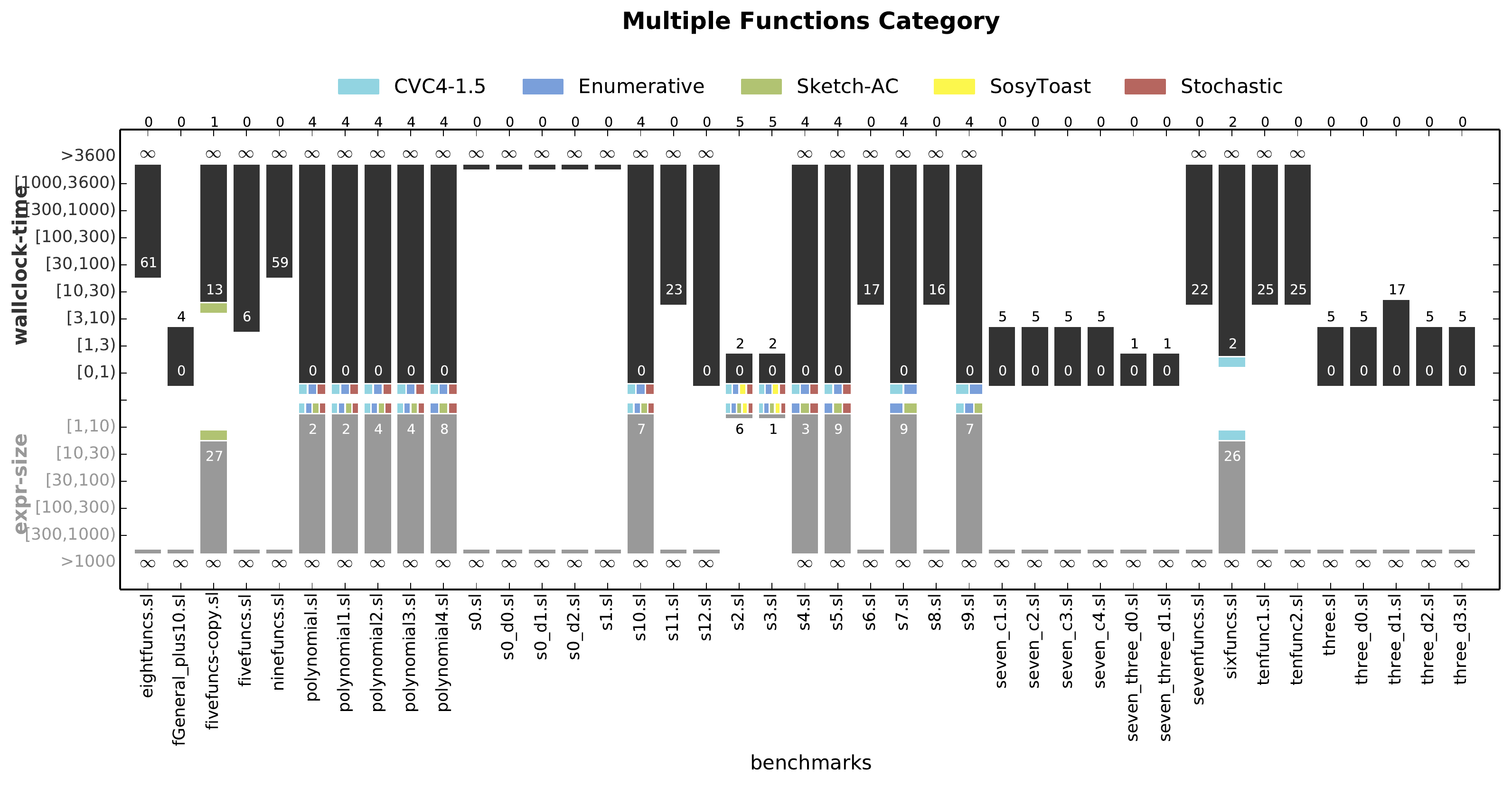} \\
\end{tabular}
}}
\caption{Evaluation of the Multiple Functions benchmarks.}\label{fig:mpmp-results}
\end{figure*}

Figures~\ref{fig:inv-results}-\ref{fig:mpmp-results} presents the results of each benchmark in the competition. The results are organized per category. The results for ICFP category are not included, since only one benchmark was solved. This benchmark is the \texttt{icfp\_28\_10.sl} benchmark, it was solved by \cvc\ and \toast, in both configurations. The two solvers (in the overall three configurations) terminated in less than two seconds, producing an expression of size two. It seems that the solvers find it difficult to handle the benchmarks in this category since the specification constraints are too partial.

\begin{figure*}
\noindent\makebox[\textwidth]{
\scalebox{0.6}{
\begin{tabular}{c}
\includegraphics[width=9.5in]{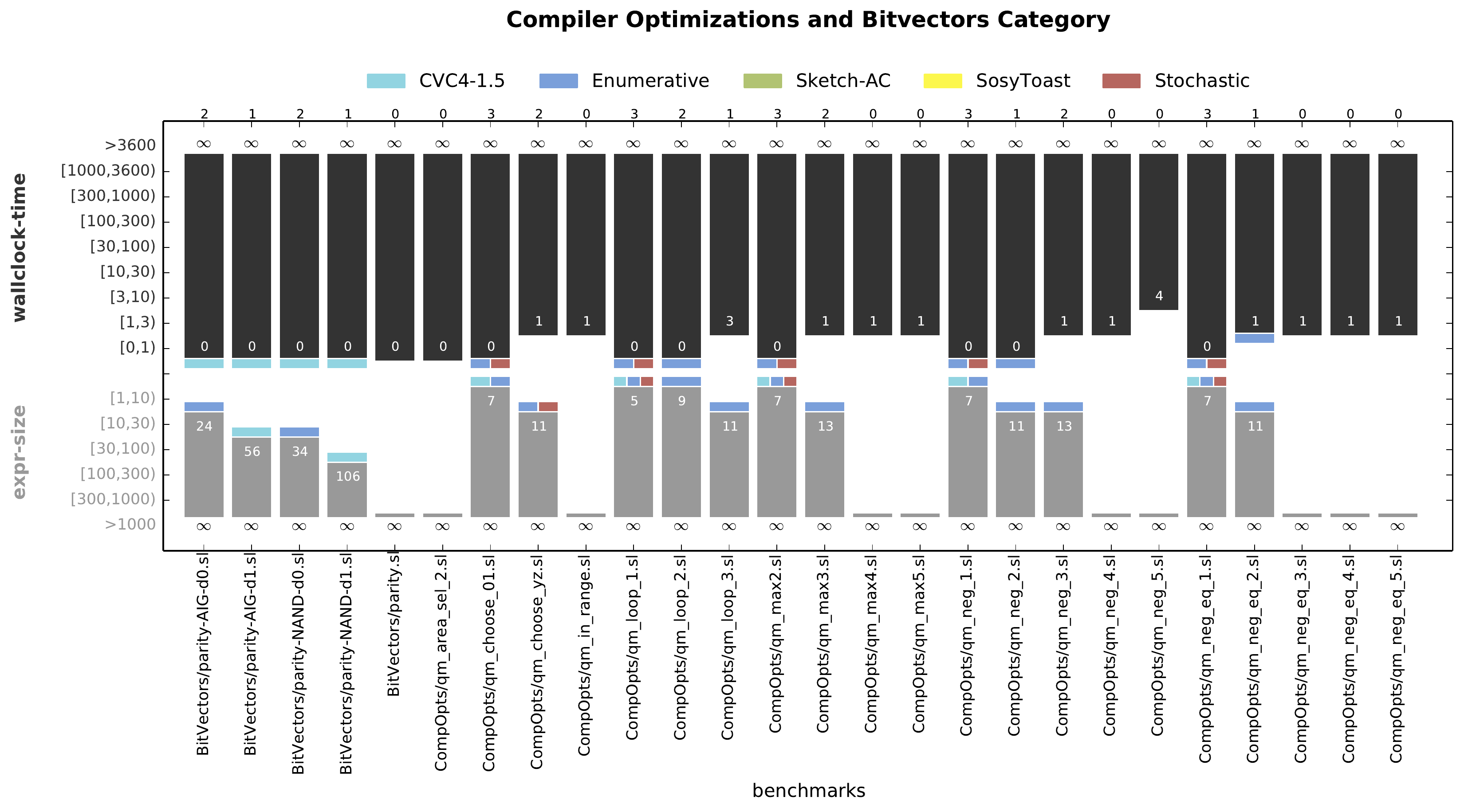} \\
\end{tabular}
}}
\caption{Evaluation of Compiler Optimization and Bitvectors benchmarks.}\label{fig:mix-results}
\end{figure*}

\begin{figure*}
\noindent\makebox[\textwidth]{
\scalebox{0.6}{
\begin{tabular}{c}
\includegraphics[width=9.5in]{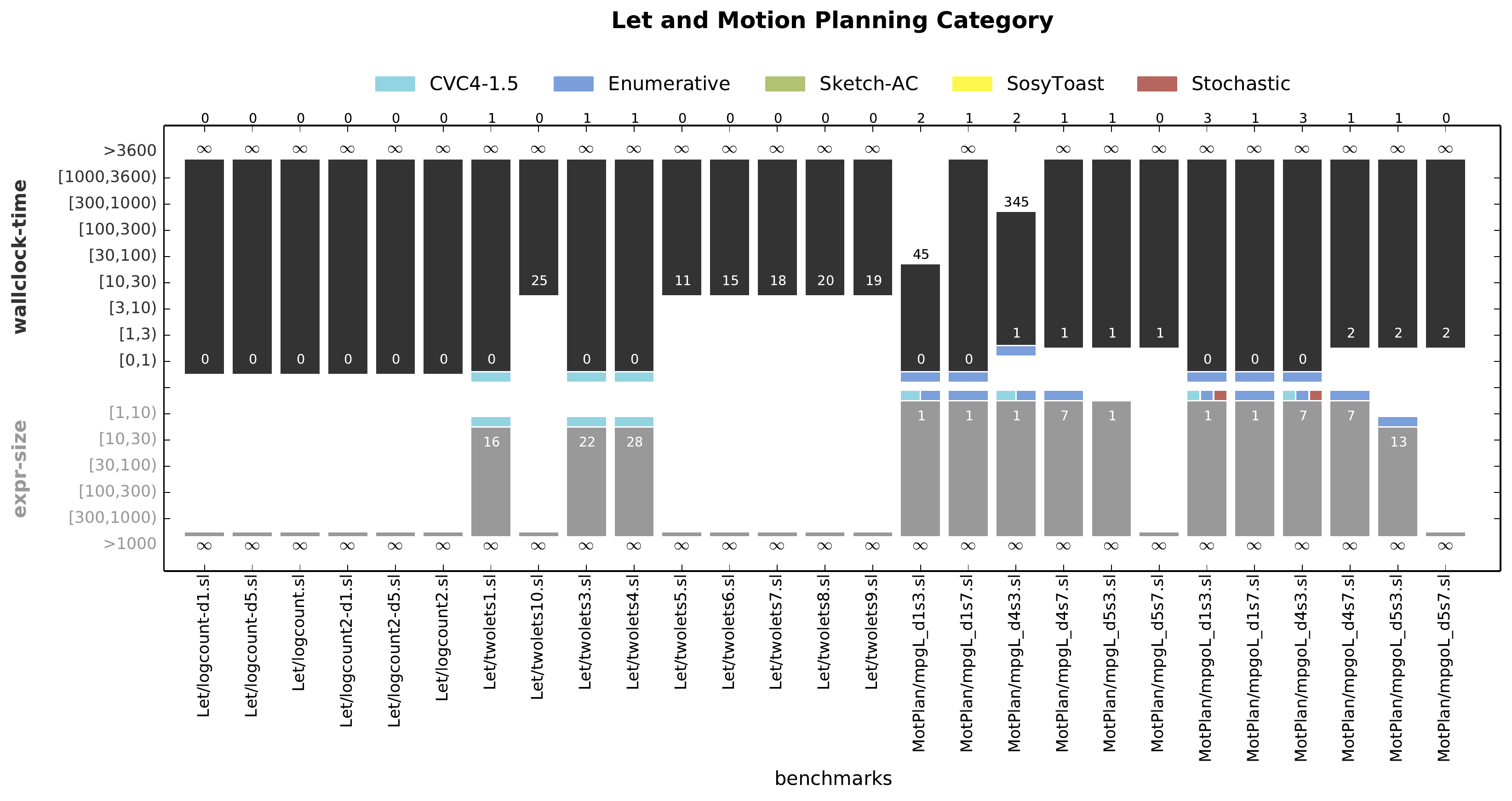} \\
\end{tabular}
}}
\caption{Evaluation of Let and Motion Planning benchmarks.}\label{fig:mix-results}
\end{figure*}

\subsection{Observations} 
\label{subsec:obs}
Correlating the number of instances solved by the different solvers (as depicted in Fig.~\ref{fig:solver-results}) and the features a certain category has (as detailed in Table.~\ref{tbl:features}) the following observations can be made:
\begin{itemize}
\item The \cvc\ solver preforms best on benchmarks with single invocation and a single unknown.
\item The \ice\ solver, designated to solve invariant synthesis problems, does best on invariant synthesis problems, which are multiple invocation.
\item The \enum\ and \stoch\ solver preform well on multiple invocation problems (of the general track). 
\item The \skac\ solver preforms well on benchmarks with a restricted grammar.
\end{itemize}

We note that we have tested the solvers submitted to the general track on the instances of the INV and LIA track translated to general SyGuS. None of them solved any of those instances. For most of the solvers the problem seemed to be lack of support for arbitrary constants --- the translation to general SyGuS included the grammar term \texttt{(Constant Int)} which allows any integer to appear in the specification, and the solvers for the general track that mostly build on guiding the search according to the syntactic structure do not support this construct. The reason \cvc\ failed on these benchmarks is that it wasn't designed to support the less common Boolean connectives such as \texttt{nand} and \texttt{xnor} that the translation to general SyGuS allowed in order to be as permissive as possible. For next year's competition we plan to test the general solvers on a restricted, yet general, subset of LIA, e.g. one consisting of just zero and one as constants and a small subset of the Boolean operators from which the others can be derived (e.g. \texttt{and}, \texttt{or}, and \texttt{not}).

With regard to size expressions we note that the \enum\ solver by definition always produces the smallest expression. Inspecting figures~\ref{fig:inv-results}-\ref{fig:mpmp-results} we observe that in many cases most solvers produced expressions of relatively the same size.  Though the \cvc\ solver often produces very large expressions. For instance, on the \texttt{array\_search\_7.s} benchmark, in which \skac\ produced an expression of size 31, \cvc's expression was of size  3,311, and on the \texttt{LinExpr\_inv1\_ex.sl}, in which both \enum\ and \skac\ produced an expression of size 14, \stoch\ produced an expression of size 42, \cvc\ produced an expression of size 4,022.
The largest expression produced was for benchmark \texttt{array\_search\_15.sl}. Its size is 1,843,271.  It is the solution of \cvc\ which is the only solver to solve this benchmark. 

\section{Discussion}
\label{sec:discussion}
The syntax-guided synthesis competition of this year (\comp'15) added two special tracks, above the general track of last year's competition: The LIA track where the background theory is conditional linear integer arithmetic and no grammar is given,  instead the grammar is implicitly assumed to be that of the LIA logic of SMT-LIB; and the INV track (also using LIA and no syntactic constraints)  consisting of  invariant synthesis problems and using special constructs that convey the structure of an invariant synthesis problem. 

A total of eight solvers were submitted to this year's competition. On the LIA and INV tracks competed three solvers (with an overlap of two), and in the general track competed five solvers in seven configurations. 
The winner of the LIA track and general track was the \cvc\ solver~\cite{ReynoldsDKTB15}, which is the only solver to participate in all tracks and the first solver to implement a syntax-guided synthesis engine inside an SMT solver. The winner of the INV track was the \ice\ solver~\cite{ICEDT}, a special solver for invariant synthesis problems.

We classified the benchmarks participating in the competition into several categories and analyzed the features a certain category has or has not. Correlating the results with the features we observe that \cvc\ preforms best on benchmarks where there is a single function to be synthesized, and that function is always invoked using the same set of parameters. On benchmarks with multiple functions to synthesize or in which functions are invoked with different parameters, the \enum\ and \stoch\ solver (which won first and second place in last year's competition) seem to preform better.

We hope this report sheds some light on the correlations of solver's performance and benchmark features, and can be of help to improve the development of future solvers.  In next year's competition we are considering devoting a track to benchmarks where obtaining a small expression is more crucial than obtaining it fast, as is the case for instance, in compiler optimization problems.

\subsection*{Acknowledgments}
We would like to thank the following people for various interesting discussions related to the competition, its tracks, the SyGuS format and various other topics related to syntax-guided synthesis:
Sarah Chasins, 
Ruediger Ehlers, 
Pranav Garg, 
Viktor Kuncak,
P. Madhusudan, 
Ken McMillan,
Daniel Neider,
Nissim Ofek, 
Arjun Radhakrishna,     
Mukund Raghothama,
Andrew Reynolds,
Heinz Riener, 
Shambwaditya Saha and
Abhishek Udupa.

We would like to thanks the StarExec team, and especially Aaron Stump, for allowing us to use their platform and for their remarkable support for \comp's special needs.  

This research was supported by US NSF grant
CCF-1138996 (ExCAPE).

\nocite{*}
\bibliographystyle{eptcs}
\bibliography{bib}
\end{document}